%% file: main.tex
\documentclass[letterpaper, 12 pt, conference]{ieeeconf}  % Comment this line out
\IEEEoverridecommandlockouts                              % This command is only
\usepackage[utf8]{inputenc}
\usepackage[T1]{fontenc}

\usepackage{amsmath,amssymb,amsfonts}
\usepackage{algorithmic}
\usepackage{algorithm}
\newsavebox{\ieeealgbox}

\usepackage{subcaption}
\usepackage{setspace}
\usepackage{graphicx}
\usepackage{textcomp}
\usepackage{xcolor}
\usepackage{float}
\usepackage{multirow}
\usepackage{url}
\usepackage{hyperref}
\usepackage{soul}
\usepackage{booktabs,makecell, tabularx}
\usepackage{nccmath}
\usepackage{pifont}
\usepackage{wrapfig}
\usepackage{listings}
\usepackage{framed}
\usepackage{needspace}
\usepackage{xcolor}         
\usepackage{adjustbox}       
\usepackage{capt-of}
\title{\LARGE \bf
LLM-Based Community Surveys for Operational Decision Making in Interconnected Utility Infrastructures
}

\author{Adaeze Okeukwu-Ogbonnaya$^{1}$, Rahul Amatapu$^{2}$,  Jason Bergtold$^{3}$ and  George Amariucai$^{4}$% <-this % stops a space
\thanks{}% <-this % stops a space
\thanks{$^{1}$Adaeze Okeukwu-Ogbonnaya, Department of Computer Science, Kansas State University}%
\thanks{$^{2}$Rahul Amatapu, Department of Computer Science, University of Kansas}%
\thanks{$^{3}$Jason Bergtold, Department of Agricultural Economics, Kansas State University}%
\thanks{$^{4}$George Amariucai, Department of Computer Science, Kansas State University}%
}

\begin{document}

\maketitle
\thispagestyle{empty}
\pagestyle{empty}

%%%%%%%%%%%%%%%%%%%%%%%%%%%%%%%%%%%%%%%%%%%%%%%%%%%%%%%%%%%%%%%%%%%%%%%%%%%%%%%%
\begin{abstract}

 We represent interdependent infrastructure systems and communities alike with a hetero-functional graph (HFG) that encodes the dependencies between functionalities. This graph naturally imposes a partial order of functionalities that can inform the sequence of repair decisions to be made during a disaster across affected communities. However, using such technical criteria alone provides limited guidance at the point where the functionalities directly impact the communities, since these can be repaired in any order without violating the system constraints. To address this gap and improve resilience, we integrate community preferences to refine this partial order from the HFG into a total order. Our strategy involves getting the communities' opinions on their preferred sequence for repair crews to address infrastructure issues, considering potential constraints on resources. Due to the delay and cost associated with real-world survey data,  we utilize a Large Language Model (LLM) as a proxy survey tool. We use the LLM to craft distinct personas representing individuals, each with varied disaster experiences. We construct diverse disaster scenarios, and each simulated persona provides input on prioritizing infrastructure repair needs across various communities. Finally, we apply learning algorithms to generate a global order based on the aggregated responses from these LLM-generated personas.

\end{abstract}


%%%%%%%%%%%%%%%%%%%%%%%%%%%%%%%%%%%%%%%%%%%%%%%%%%%%%%%%%%%%%%%%%%%%%%%%%%%%%%%%
\section{Introduction}
Natural disasters often damage multiple critical infrastructures across communities, which creates complex prioritization challenges for response efforts.  Traditional approaches for prioritizing infrastructure repairs rely heavily on expert opinions, which potentially overlook diverse community perspectives and public demand urgency.  A recent study by Doorn et al \cite{doorn2019multidisciplinary} shows that restoring damaged infrastructure during disasters following the technical approach is likely ineffective in addressing the immediate needs of affected communities. Also, studies have shown that vulnerable communities have higher immediate needs and are most likely to be affected if such needs are not met \cite{eisenman2007disaster}\cite{paton2001disasters}.

Due to this, the current research trend is focused on integrating the social needs of communities into the technical aspects of infrastructure restoration to achieve a fair allocation of limited resources during disasters \cite{zhai2020examine}\cite{drakes2021social}\cite{domingue2019social}\cite{chen2021public}. Building upon these efforts, this paper proposes additional prioritization schemes that incorporate community perspectives, thereby complementing existing approaches and enhancing the overall fairness and effectiveness of the infrastructure recovery process. We leverage the use of Large Language Models (LLM) to simulate diverse community opinions and generate synthetic data on the community's preferred orders of repair. Afterward, we use a ranking algorithm to develop a total order for prioritizing infrastructure repairs in communities.  

While fairness may often be treated as a technical goal, for instance, like ensuring access to all or prioritizing the most vulnerable, it is also a socially constructed and context-dependent concept. What a community may perceive as fair may differ based on their values, lived experiences, or crisis-specific needs. In 
using LLM, we heavily rely on our prompt design to encode assumptions about fairness, urgency, or moral trade-offs. This might reflect in our results, such that prompts that emphasize vulnerability may lead to prioritizing high-risk areas, and prompts that focus on service restoration might focus on high-density or populated areas. This means that the fairness in our results is not an objective truth rather a reflection of the values embedded in our prompts. Hence, we acknowledge that prompt design may influence what the model learns to treat as fair.

This study serves as proof of concept in incorporating community preferences and priorities into recovery and restoration after a natural disaster or hazard, demonstrating how LLM-generated preferences can complement infrastructure prioritization models. In future work, we plan to validate this approach using real community surveys or explore using prior survey data to generate synthetic preferences for new contexts, particularly where survey response rates are low. Our contributions include the following:
\begin{enumerate}
    \item We introduce a hetero-functional graph of a toy infrastructure model comprising power, water, and community infrastructures . We use this graph to establish a partial order showing the sequence of repair decisions to be made during disaster or post-disaster across affected communities.
    
    \item We use a Large Language Model (LLM) as a survey tool and obtain community preferences as ordered pairs. 

    \item We use a neural network-based ranking algorithm to obtain a total order of repair prioritization, which is a consensus of the community preferences.
    
\end{enumerate}

The rest of this paper is organized as follows: Section 2 reviews related work on the use of hetero-functional graphs, infrastructure repair prioritization, the use of LLM in surveys, and order learning. Section 3 describes the proposed methodology, including the LLM-based simulated community preferences, the comparator and chainization model, random subset sampling, and the sensitivity analysis. Section 4 describes the data collected during the experimentation. Section 5 presents experimental results and key findings. Finally, in Sections 6 and 7, we discuss the findings in our research and conclude the paper, outlining directions for future research.

\section{Literature Review}
Hetero-functional graphs (HFGs) have been employed to model complex interdependent infrastructure systems, particularly in the context of smart cities and utility networks \cite{schoonenberg2019hetero}\cite{thompson2024hetero}\cite{munikoti2021robustness}. For instance, Schoonenberg et al. \cite{schoonenberg2019hetero} developed a comprehensive HFG model to represent interdependent smart city infrastructures, demonstrating its applicability in capturing the multifaceted interactions among urban systems. Similarly, Munikoti et al. \cite{munikoti2021robustness}  utilized HFGs to assess the robustness of interdependent urban utility networks, providing insights into system vulnerabilities under various failure scenarios. Despite these advancements, the application of HFGs in disaster recovery, specifically for prioritizing infrastructure repair, remains unexplored. Our work extends HFG modeling into this domain, applying it as a tool for extracting partial orders of functionality during disaster events. This allows us to formalize system-level dependencies and provide structure to otherwise ambiguous recovery decisions.

Recent research has explored how Large Language Models can be used in persona generation and generating opinions in surveys or stand-ins for community feedback. This approach becomes more important, especially when real-time or large-scale surveys are difficult to execute. These studies lay the groundwork for synthetic survey and preference generation using LLMs. Dolant and Kumar \cite{dolant2025agentic} proposed a multi-agent LLM in which simulated personas assume distinct stakeholder roles to support decision discourse under disaster conditions. They explored tradeoffs in decision-making and in strategy generation by using a flood scenario in a MidWestern township where personas exhibit evolving priorities. Li et al. \cite{li2025llm} conducted large-scale empirical evaluations of LLM-generated personas for simulating public feedback, including in use cases such as U.S. election forecasting. Their study critically identified systematic biases and deviations from real-world data, which they attribute to ad hoc generation techniques. Shi et al. \cite{shi2024argumentative} used LLM-generated personas in structured multi-perspective debates to simulate community perspectives on complex topics. Though not explicitly survey-based, the study showed how synthesized viewpoints can approximate population-level opinion diversity and influence the ranking of decisions by exposing users to multiple attitudes and concerns. These papers demonstrate the potential of LLMs to produce community inputs through simulations that can substitute for or augment surveys. 

Xie et al \cite{xie2025ragbasedmultiagentllmnatural} presented WildfireGPT, a multi-agent LLM system that tailors decision support to different stakeholders during wildfire scenarios. In this study, community-based feedback is used to inform risk reports via retrieval-augmented LLMs. Alqitham \cite{alqithami2025integrating} introduced a system that uses real-time public sentiment from social media along with reinforcement learning to improve equity in disaster resource distribution. In this work, community feedback is ranked and weighted using optimization to guide the allocation of utility resources such as water, electricity, and medical aid. Chen et al. \cite{chen2024intelligent} designed a decision-support system that combines LLMs with multi-objective optimization algorithms and reinforcement learning for dispatch and resource allocation in urban emergencies. In this work, user feedback is collected iteratively and used for personalized allocation in urban emergency response. Chen et al. \cite{chen2024enhancing} built a structured emergency system using knowledge graphs to support LLM reasoning during crises. Though these works are not focused on graphs of interdependent infrastructures, they applied LLM-based systems using community feedback to guide real-time or resource allocation decisions in disaster management. 

As LLMs are used more often to simulate people, concerns about reliability and fairness have increased. Li et al. \cite{li2025llm} pointed out that ad hoc persona generation can introduce systemic biases. They argue that a lack of control procedures can cause downstream distortions, especially in models meant for decision support. Shi et al. \cite{shi2024argumentative} tested whether synthetic views influence human judgment by tracking eye movements and engagement metrics. The study raises critical concerns about the power of synthetic viewpoints to shape judgment. 

These studies emphasize the need to evaluate synthetic feedback carefully. Our approach addresses this by performing sensitivity analyses by varying the prompts to the LLM. This allows us to detect and explain potential preference bias in the LLM-generated rankings.

Recent works show that infrastructure repair and investment prioritization rely on both technical and social criteria. Technical factors are usually based on asset condition, efficiency, cost-benefit, and performance objectives, while social factors include equity, social vulnerability, and stakeholder input \cite{mohamadiazar2024equitable}\cite{hastak2005decision}\cite{das2023multi}\cite{garfi2011decision}\cite{dell2024multicriteria}\cite{marcelo2016prioritizing}\cite{dhansinghani2022design}\cite{karasneh2024priority}\cite{galal2013integrating}. These solutions often use tools like multi-criteria decision analysis (MCDA) and analytic hierarchy process (AHP) to combine different types of criteria into a final score or ranking.

Even though social criteria are included, technical factors tend to carry the most weight and usually drive the final decision. Pramesti et al. \cite{pramesti2021bridge} show that while economic and social factors are considered, they are limited in scope and mainly reflect the views of stakeholders and regulators. Karasneh and Moqbel \cite{karasneh2024priority} find that technical factors dominate, while the influence of socioeconomic and environmental factors is minor. Das and Nakano \cite{das2023multi} operationalize social impact metrics alongside technical ones, but actual community participation is not included. Bauer et al. \cite{baur2003multi} show that although customer satisfaction and external benefits are considered, community preferences are not directly built into the decision-making process for the annual rehabilitation programs in drinking water networks. The focus remains on technical performance and service delivery.

Some works show that including social aspects can shift the rankings when done well. The study by Mohamadiazar et al. \cite{mohamadiazar2024equitable} shows that incorporating social and hazard vulnerability changes which bridges are seen as most critical. In Dhansinghani et al. \cite{dhansinghani2022design}, both stakeholder input and technical criteria are used. One result of stakeholder involvement was that separating cost from the priority score gave more weight to social goals like walkability and school access. Dell’Anna et al. \cite{dell2024multicriteria} also show that adding vulnerability-related criteria can re-rank urban regeneration projects compared to a technical-only approach. Although they do not collect community surveys, they still reflect equity concerns through weighted criteria.

Although many prioritization methods include social criteria, most do not gather or use community preferences to guide final decisions. We observe that technical scores still take the lead in most cases. In some works where social criteria are weighted carefully or when stakeholder input is added, it affects the priorities being studied.  In our work, we focus on cases where there is no technical total order, which is when functionalities can be repaired in any order. We begin with a technical prioritization, which is the partial order provided by the hetero-functional graph, and resolve the remaining ambiguity by including community preferences from LLM simulations. 

Recent studies in order learning focus on how to create a full ranking from a limited number of pairwise comparisons. Some of these approaches treat the problem like matrix completion, where missing preferences are filled in based on patterns in the data. For example, Park et al. \cite{park2015preference} model preference data as a low-rank score matrix and show that a small number of comparisons can still give good results. Gunasekar et al. \cite{gunasekar2016preference} go further by recovering the full ranking from a partial order (represented as a directed graph) without needing exact scores, using a method called nuclear-norm regularization.

Other researchers use semi-supervised learning, which combines labeled and unlabeled data. Szummer and Yilmaz \cite{szummer2011semi} add a graph-based regularizer that ensures similar items get similar ranks. More recently, neural network models have been applied to this problem. He et al. \cite{he2022gnnrank} propose GNNRank, a graph neural network that learns from comparison graphs and penalizes incorrect rankings, achieving strong results even with limited data. Lee and Kim \cite{lee2022order} introduced the chainization method, which builds a full ranking by starting with a partial order and adding new “pseudo” comparisons to improve training.

We build on these ideas by adapting chainization \cite{lee2022order} to infrastructure planning. We start with pairwise preferences generated by LLMs, train a model to compare options, and then use pseudo-pairs to improve the ranking. This helps us build a complete repair priority list even when we cannot collect all possible comparisons from the community.

\section{Methodology}
In this section, we discuss the approach of modeling infrastructure dependencies and generating a partial order using a hetero-functional graph representation. Next, we discuss the use of Large Language Models (LLMs) to generate community preferences. We describe the aggregation of these preferences and, finally, the machine learning-based chainization approach for ordering functionalities to produce a total repair order.

\subsection{Hetero-Functional Graph (HFG) Model}
We use a hetero-functional graph to represent the infrastructure system and its interdependencies. An HFG allows us to model different types of functionalities, such as generation, transportation, storage, and consumption of operands like power, water, and vehicles. Each functionality is represented as a node in the graph, and directed edges capture the dependency relationships between them. 

\begin{figure}[ht]
    \centering
    \includegraphics[width=9cm]{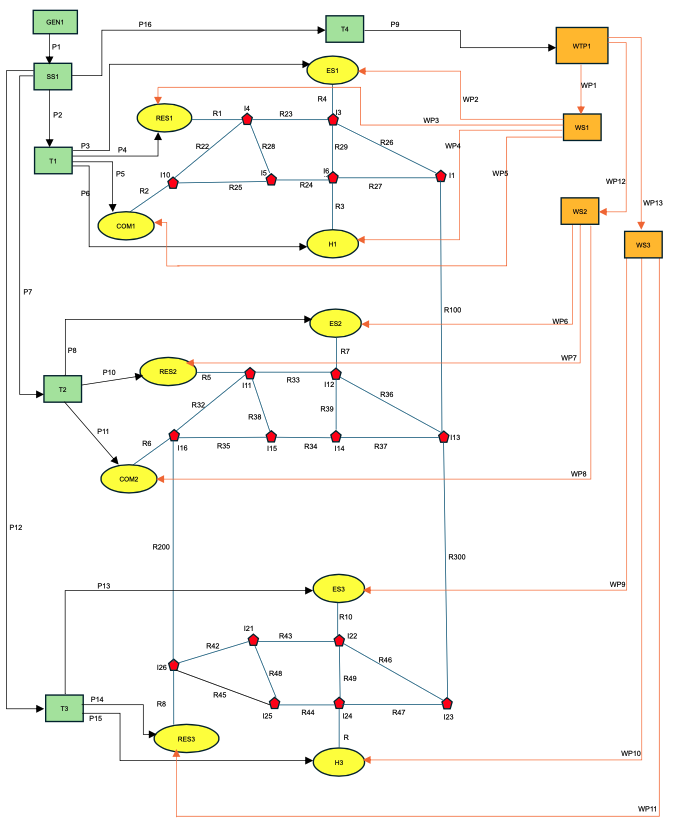}
    \caption{The Integrated Infrastructure and Community Network Model}
    \label{fig:model1}
\end{figure}

The HFG includes infrastructure functionalities across multiple communities. These communities are linked to the infrastructure nodes through consumption and service-related functionalities. We use this structure to capture both physical infrastructure flow and its connections to community-level services. Figure~\ref{fig:model1} shows three distinct communities we use in this study, each featuring dedicated subsystems for electrical power distribution and water supply. The hetero-functional graph used in this study is publicly available at \cite{Okeukwu2025_rank}. 

\begin{figure*}[h]
    \centering
    \includegraphics[width=1\textwidth]{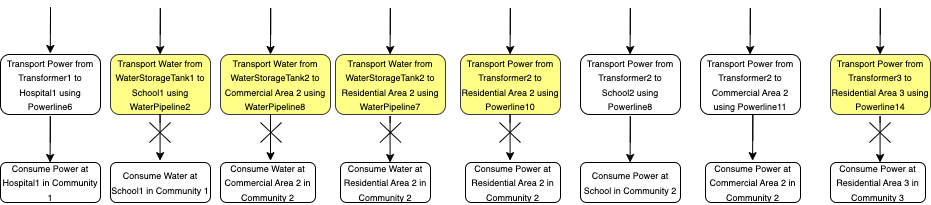}
    \caption{A subset of the hetero-functional graph (HFG) showing dependencies across power and water systems. The final layer consists of consumption functionalities that can be repaired in any order. Arrows represent functional dependencies.The highlighted functionalities are damaged while the links marked $\times$ are paths requiring repair before the associated community-level services (bottom layer) can be restored.}
    \label{fig:hfg_diagram}
\end{figure*}

From the HFG, we extract a partial order of functionalities. This partial order tells us which functionalities must be repaired before others based on system dependencies. A total order, on the other hand, is a complete ranking of all functionalities using some metric or preference criteria. In the final layer of the graph, where consumption nodes like “Consume Water in Residential Area 1” or “Consume Power in Residential Area 2” are located, we notice that many of these do not depend directly on each other. That means they can technically be repaired in any order as long as their upstream dependencies are already working. Figure~\ref{fig:hfg_diagram} shows the bottom part
of the HFG used in our model. The final layer contains functionalities directly tied to community-level services, where technical prioritization is not sufficient.

In our study, we focus on the fact that consumption functionalities may stop working at the time of disaster because their associated upstream functionality, such as transport power to Residential Area 2 using PowerlineX or transport water to Commercial Area 1 using WaterPipelineX, may be damaged. Therefore, the consumption functionalities depend on repairs being made to their upstream dependencies before they come on. Once those dependencies are fixed, the consumption nodes become available. Since consumption functionalities do not depend on one another, the model gives no guidance on which upstream dependency should be repaired first.

This creates ambiguity when resources like repair crews are limited. The technical model tells us what is possible and necessary, but it does not tell us what should take priority when multiple actions are equally feasible. To solve this, we turn to community preferences. In the next stage of our method, we bring in input from simulated personas to help resolve this ambiguity.

\subsection{LLM-based Simulated Community Preferences}
We simulate community input and generate responses from synthetic personas using large language models (LLMs). The personas we create represent a diverse set of individuals from different communities with varied backgrounds, social contexts, and disaster experience. Our goal is to approximate how people in communities might prioritize infrastructure repairs during a disaster when multiple repair actions are possible.

\subsubsection{Persona Generation}
We generate 200 personas by applying prompting using few-shot learning in LangChain. We define each persona using 44 attributes from the American Census Survey(ACS), such as age, occupation, household composition, and several others. We add some additional attributes, such as disaster experience and geographical class. The full list of attributes is in Table~\ref{fig:persona_attributes}.

We use the GPT-3.5-turbo model with a temperature of 0.9. All personas were generated on April 2nd, 2025, between 2:00 PM and 4:00 PM Central Time using the OpenAI API. 

We use two persona datasets: one where each persona is assigned to a community (\emph{community-aware}) and another where community identity is removed (\emph{community-unaware}). Both share the same attributes. We generate the community-aware set first and then create the community-unaware version by removing community labels. This allows us to test whether community awareness affects persona preferences. See generated persona data here \cite{Okeukwu2025_rank}.

\begin{table}[ht]
\centering
\scriptsize
\setlength{\tabcolsep}{2.5pt}
\caption{Attributes Used to Generate Synthetic Personas}
\begin{tabular}{ll}
\hline
\textbf{Attribute} & \textbf{Description} \\
\hline
Persona ID & Unique identifier for each persona \\
Age & Age in years \\
Gender & Gender identity \\
Race/Ethnicity & Self-identified racial or ethnic group \\
Marital Status & Marital or relationship status \\
Citizenship Status & Citizenship classification \\
Birth Place & Country or region of birth \\
Year of Immigration & Year of immigration (if applicable) \\
Primary Language & Main language spoken \\
English Proficiency & Level of fluency in English \\
Education Level & Highest level of education attained \\
School Enrolment & Current enrolment status \\
Employment Status & Employment situation \\
Occupation & Job or role in the workforce \\
Industry of Employment & Sector or field of employment \\
Class of Worker & Employment classification \\
Work Hours Per Week & Average hours worked weekly \\
Income Bracket & Annual income range \\
Poverty Status & Position relative to poverty line \\
Government Assistance & Participation in support programs \\
Household Relationship & Role in household (e.g., head, dependent) \\
Household Size & Number of people in the household \\
Number of Children & Number of children in the household \\
Responsible for Grandchildren & Caretaking responsibility \\
Home Ownership & Own or rent housing \\
Housing Type & Type of dwelling \\
Utilities Access & Availability of basic utilities \\
Utilities Housing Costs & Cost of utilities and housing \\
Transportation Mode & Main mode of transport \\
Commute Duration & Travel time to work \\
Work From Home & Whether the persona works remotely \\
Health Insurance & Type of health coverage \\
Disability Status & Whether the persona has a disability \\
Disability Type & Type of disability (if applicable) \\
Military Service & Military background \\
Military Service Period & Time served in the military \\
Disability Rating & Government-assigned disability rating \\
Internet Access & Type of internet connection at home \\
Device Access & Devices available to the persona \\
Recent Birth & Whether persona recently had a child \\
Moved in Last Year & Whether persona changed residence \\
Previous Residence & Where persona lived before moving \\
Disaster Experience & Experience with past disasters \\
Community Description & Infrastructure status across 3 communities \\
Geographical Class & Urban, suburban, or rural location \\
\textbf{Community*} & Community assignment \\
\hline
\end{tabular}
\label{fig:persona_attributes}
\end{table}

\subsubsection{Simulation Setup}
We assign each persona a disaster scenario with infrastructure disruptions across three communities. For each scenario, we specify which infrastructure functionalities are working and which are damaged. The damaged functionalities are selected to create a mix of affected services and sectors, requiring personas to choose between, for example, water access in residential areas and power in commercial areas.

The information provided about each community includes the list of working and damaged functionalities, its geographical classification (urban, suburban, or rural), and its social vulnerability score. For example, Community 1 is classified as rural and has a social vulnerability score of 2.0 out of 10. In this community, water in the school is damaged, and both water and power in the residential area are down, while the hospital and commercial area remain functional. Community 2 is suburban with a vulnerability score of 3.0. Here, the residential area and water access in the commercial area are damaged, while the school and other infrastructures are operational. Community 3 is urban with a high vulnerability score of 7.0. In this case, water access is disrupted in school and residential areas, and power is also down in the residential area, while the rest of the infrastructure is functional.

We include this information in the prompt so that the persona can consider both system-wide infrastructure conditions and the community context when making decisions. This setup allows us to simulate trade-offs in prioritization.

\subsubsection{Question and Response Generation}
For questions, we create all possible pairwise comparison questions by taking all unique pairs of damaged functionalities across the three communities. To collect repair preferences, each persona is presented with a disaster scenario where multiple infrastructure functionalities are damaged. Then, the persona is asked to choose between two feasible repair options and to provide a justification for their choice. Each prompt is designed to capture both the decision and the reasoning behind it. In every prompt, we make sure the persona has access to information about the status of all three communities. We repeat this process across many personas and question combinations to cover all pairwise preferences we created. We use the GPT-3.5-turbo model with a temperature of 0.7.  We did these generations on April 3rd, 2025, between 2:00 PM and 4:00 PM Central Time using the OpenAI API. Previously, we mentioned that we used a temperature of 0.9 in the persona generation. This was to promote diversity in characteristics such as age, occupation, and disaster experience. Now, for the response generation, we set the temperature to 0.7 to encourage more stable, realistic, and context-grounded reasoning. With this configuration, we try to balance creative variability during persona creation with more deterministic and coherent outputs during response collection. The specific prompt templates used to generate the personas and collect their responses are provided in the Appendix.

\subsection{Feature Extraction \& Data Processing}
We group the data by question and calculate the percentage of personas who chose each option. These percentages serve as soft labels, a two-dimensional probability vector $[p_1,p_2]$ where $p_1$ is the number of personas who chose option 1 and $p_2$ is 1-$p_1$. For each pairwise question, the dataset is a pair of feature vectors and the corresponding soft label.

Next, we extract features for each repair option, which represent damaged functionalities in specific communities. Each feature vector includes information about the infrastructure type and its community. Table~\ref{fig:features} lists the features we use for each option. The SVS\cite{enderami2024social} is a metric we use to identify the different vulnerability levels across a community during disasters, originally defined on a 1–5 zone scale. In our work, we normalize this to a 0–1 scale, with 1 representing the highest vulnerability. In total, we aggregate 36 distinct pairwise questions drawn from 7,200 persona responses.

\begin{table}[ht]
\centering
\scriptsize
\setlength{\tabcolsep}{2.0pt}
\caption{Infrastructure Repair Option Features}
\label{fig:features}
\begin{tabular}{ll}
\hline
\textbf{Feature} & \textbf{Description} \\
\hline
Infrastructure Type & One-hot encoded: Water, Power \\
Facility Type & One-hot encoded: Hospital, Residential,\\ &Commercial, School \\
Community ID & One-hot encoded: Community 1, 2, or 3 \\
Geographical Class & One-hot Encoded: Urban, Suburban, \\ &or Rural \\
Social Vulnerability Score (SVS) & Continuous value (0 to 10) \\
\hline
\end{tabular}
\end{table}

\subsection{Pairwise Comparator and Chainization Model}
We train a neural network model to predict preferences between pairs of infrastructure repair options. The architecture consists of a feature extractor that processes each option independently, followed by a comparison layer that combines the extracted features. The output is a softmax probability distribution over the two choices.

We train the model with a hybrid loss function, where cross-entropy loss is for when the majority of personas strongly prefer one option($ P_1 \neq P_2 $), while we use KL-divergence loss when both options are equally preferred ($ P_1 = P_2 $).

\begin{figure}[ht]
  \centering
  \includegraphics[width=\linewidth]{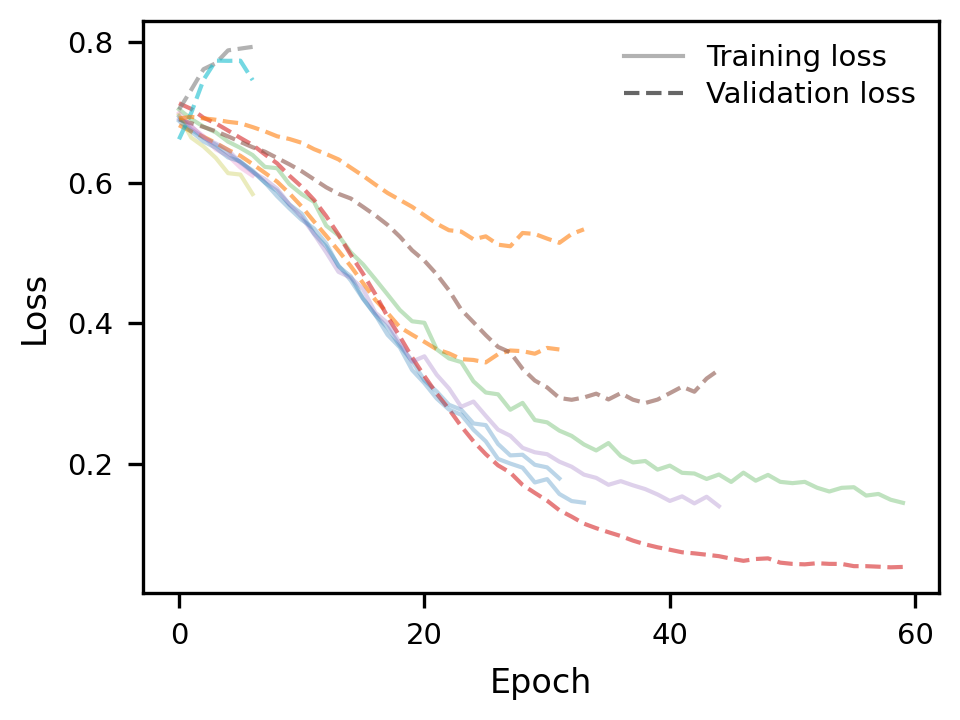}
  \caption{Six-fold cross-validation curves for the pairwise comparator at
  \(\text{LR}=1\times10^{-3}\). Solid lines show training loss; dashed lines show validation loss for each fold.}
  \label{fig:cvloss}
\end{figure}

For optimisation, we use Adam and a batch size of 8. We select the learning rate by using a six-fold validation of the full set of 36 pairwise questions. For each learning rate tested($1\times10^{-3},\,5\times10^{-4},\,1\times10^{-4}$), we obtained the best validation loss reached for every fold before early stopping. Then we averaged these six values. The mean best-fold losses were 0.4267, 0.4842, and 0.6584, respectively. We adopt the learning rate of $1\times10^{-3}$ for all subsequent experiments because that gave us the lowest average. See the training and validation curves for this setting in Figure~\ref{fig:cvloss}.

Once pairwise comparisons are learned, we construct a global ranking of infrastructure repairs using the Chainization Algorithm. To construct the pairwise preference graph, we use a directed graph, $ G = (V, E) $ where the nodes($V$) represent infrastructure repair options and edges($E$) show preference such that when option 1 is preferred over option 2 based on model prediction, an edge $(option2 \rightarrow option1)$ is added.

For the ranking, we apply PageRank instead of Kahn’s topological sorting to order infrastructure repairs. In all PageRank computations, we assign edge weights based on the model’s predicted preference confidence. We use PageRank because it handles cycles cleanly by assigning global importance scores. Then we rank infrastructure options by score to produce the final prioritization.

\subsection{Random Subset Sampling and Fine-Tuning with Pseudo-Pairs}

We adopt a training and refinement approach to investigate the feasibility of reducing survey burden while maintaining prioritization accuracy. First, we randomly sample a fraction ($20\%$, $30\%$, $40\%$, $50\%$, $60\%$, $75\%$) of the available pairwise comparisons from the full dataset. This simulates realistic scenarios where it is impractical to collect all possible preference judgments.

Then we train a pairwise comparator model using only the sampled subset. We still use the same hybrid loss formulation, that is, cross-entropy loss for strong preferences and KL-divergence loss for equal cases.

After this initial training, the comparator model predicts pairwise preferences for all infrastructure options. We use these predictions to build a directed preference graph and apply chainization to infer a preliminary global ranking.

Based on the recovered initial global rank, we generate synthetic pseudo-pairs. We do this by assuming that for any two facilities that have a clear ranking relationship (e.g., \(A > B\)), a pseudo-comparison is added with a hard label indicating definite preference,i.e., [1.0, 0.0]. We use this to enrich the available training data. Then, we fine-tune the comparator model on a combined dataset that comprises both the original sampled comparisons and the generated pseudo-pairs. Following fine-tuning, final pairwise predictions are used to reconstruct the preference graph and produce a global repair prioritization.  

We present the results of Kendall's $\tau$ correlation between the full ranking and the sampled rankings, as well as the Top-K-Lists overlap(Top-3 and Top-5)\cite{zuk2012ranking}\cite{mundra2023koios} in Section \ref{sec: results}.

\subsection{Validation and Sensitivity Analysis}
To assess the robustness of our prioritization approach, we conduct two complementary evaluations. First, we examine the influence of community identity on repair prioritization by comparing the global rankings derived from two distinct datasets: one generated using community-aware personas and the other using community-unaware personas. Second, we conduct a sensitivity analysis to test the stability of our results with respect to variations in prompt formulation. Prior studies have shown that large language models (LLMs) can be highly sensitive to prompt rewording, tone, and instruction style~\cite{mizrahi2024state}\cite{chatterjee2024posix}\cite{gandhi2025prompt}\cite{he2024does}. To evaluate this, we create six variations of the original prompt:
\begin{enumerate}
    \item Reworded Question: We changed the phrasing of the question from ``Which should be repaired first:`{option1}' or `{option2}'?'' to ``Between the two options, `{option1}' and `{option2}', which do you deem most critical for immediate repair?''.
    
    \item Reworded Reasoning: We changed the reasoning from ``Your one-sentence explanation'' to ``Explain your selection with a concise rationale, highlighting relevant factors influencing your decision.''.
    
    \item Tone Shift: We introduced empathetic framing like ``Keep in mind the emotional and practical consequences of the outage for residents living in these communities as you evaluate what should be repaired first.''~\cite{gandhi2025prompt}.
    
    \item Reformatted Instruction Style: We change the structural layout of the prompt~\cite{he2024does} by adding steps for each information in the prompt.

  \item Choice Before Reasoning: We restructured the prompt so that the LLM first states its repair choice and only then provides its justification. All other variants (including the original) require the model to explain its reasoning before revealing its selection.
  
  \item SVS Removed: We dropped the explanation of SVS to test its influence. In exploratory runs, we observed that when SVS was not defined in the prompt, the model sometimes misread the community with a low vulnerability score of 0.2 as having high vulnerability, so removing that detail helped to understand its impact.

\end{enumerate}

We use each variant to regenerate the persona responses for the same infrastructure repair comparisons. These new responses are then processed through the full modeling pipeline. To quantify robustness, we compute the Kendall $\tau$ correlation between each prompt-variant ranking and the baseline ranking from the original prompt. This allows us to determine whether minor variations in prompt design affect prioritization outcomes. Also, we look at the Top-K-Lists overlap(Top-3 and Top-5)\cite{zuk2012ranking}\cite{mundra2023koios}. All code and generated data used in this study are publicly available at the GitHub repository \cite{Okeukwu2025_rank}.

\section{Results}\label{sec: results}
We focus our results on three different aspects: the impact of community awareness in persona generation, the effect of limiting the number of preference queries using chainization and pseudo-pairs, and the sensitivity of prioritization outcomes to variations in prompt design.

\subsection{Impact of Community Awareness in Simulated Preferences}
To investigate whether awareness of community assignment affects infrastructure repair prioritization, we compare the results from community-unaware personas and community-aware personas. In the community-aware setting, personas were 
told which community they belonged to, which allowed for potential bias toward their own community. In contrast, in the community-unaware setting, personas made decisions without this information.

%Table~\ref{fig:ranking} shows the distribution of final repair rankings between community-aware and community-unaware personas. The rankings are largely consistent across both datasets, with only a minor shift in the top two priorities. In the community-unaware setting, the top-ranked repair is Water access in a school in Community 3, whereas in the community-aware setting, Residential power in Community 3 takes the top position. This subtle difference may suggest that awareness of community identity can nudge prioritization toward more individualized needs, but overall, the global preference structure remains stable even with a slight change in persona context.

Table~\ref{fig:ranking} shows the final repair ranking, which is identical for both the community-aware and community-unaware personas. We initially expected that making personas aware of their own community would introduce some level of in-group bias and shift repair priorities towards their own group. However, the consistency in rankings suggests that personas weighed other factors more heavily in their decision-making. In this scenario, community identity did not significantly influence the overall prioritization.

Also, the final ranking shows that repairs related to schools and residential areas appear most frequently near the top. Water in a school in Community 3 is ranked first, followed by  Water School in Community 1. School-related repairs appear in two of the top three positions.

Residential infrastructure fills most of the middle rankings, with power generally ranked above water within the same community. We notice that Water in Residential in Community 3 ranks above Power Residential in Community 1. This might be linked to the high vulnerability of Community 3. The only commercial repair, Water Commercial in Community 2, is ranked last.

\begin{table}[ht]
\centering
\scriptsize
\caption{Final Global Repair Prioritization (Same for Community-Aware and Community-Unaware Datasets)}
\label{fig:ranking}
\begin{tabular}{rl}
\hline
\textbf{Rank} & \textbf{Repair Option} \\
\hline
1 & Repair Water School in Community 3 \\
2 & Repair Water School in Community 1 \\
3 & Repair Power Residential in Community 2 \\
4 & Repair Power Residential in Community 3 \\
5 & Repair Water Residential in Community 3 \\
6 & Repair Power Residential in Community 1 \\
7 & Repair Water Residential in Community 2 \\
8 & Repair Water Residential in Community 1 \\
9 & Repair Water Commercial in Community 2 \\
\hline
\end{tabular}
\end{table}

\begin{figure}[ht]
    \centering
    \includegraphics[width=0.5\textwidth]{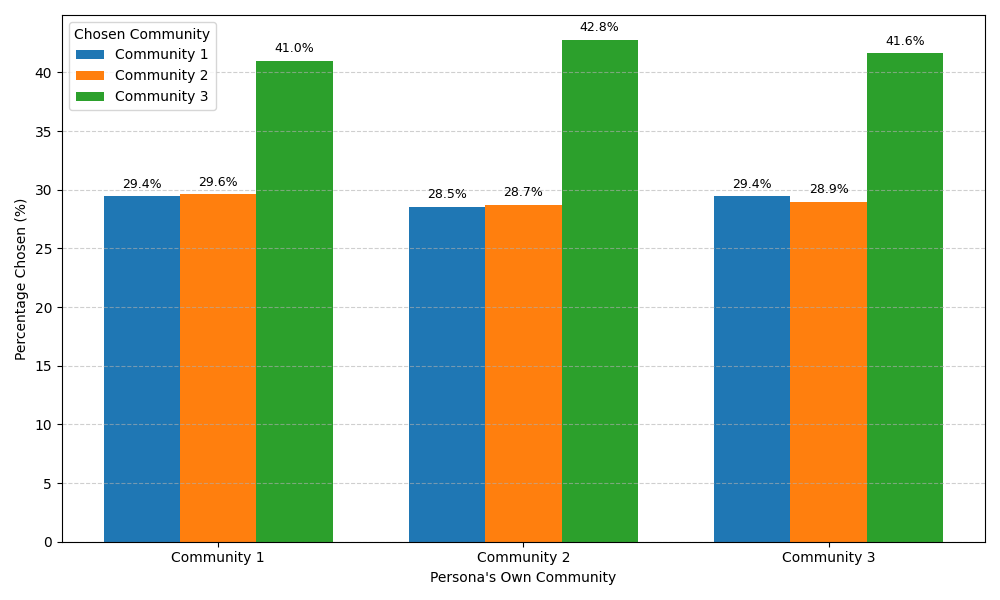}
    \caption{Community Preferences by Community-Aware Personas}
    \label{fig:comm_aware}
\end{figure}

\begin{figure}[ht]
    \centering
    \includegraphics[width=0.5\textwidth]{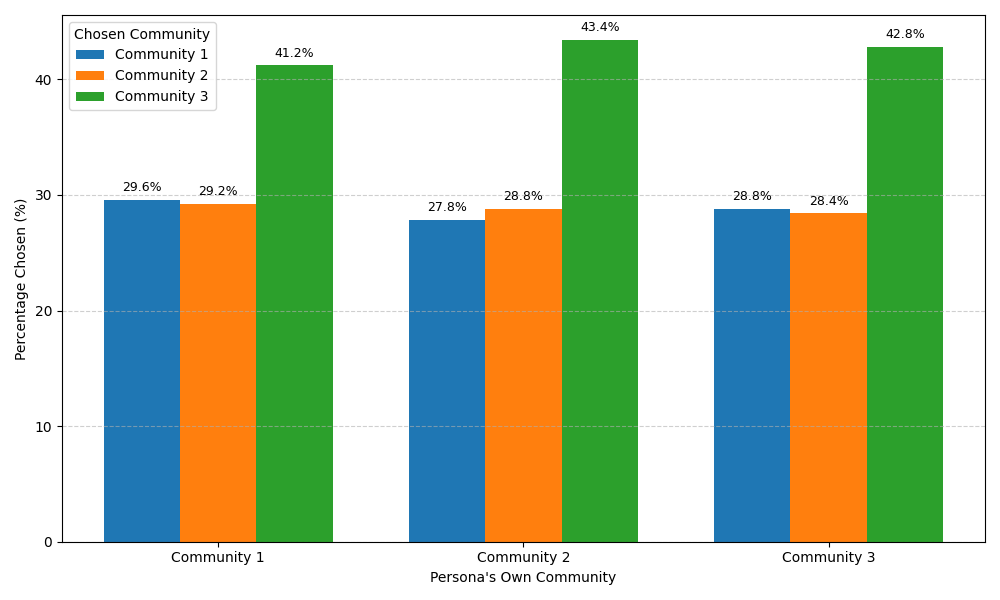}
    \caption{Community Preferences by Community-Unaware Personas}
    \label{fig:comm_unaware}
\end{figure}

Figure~\ref{fig:comm_aware} and Figure~\ref{fig:comm_unaware} show how personas from each community made choices across all three communities, in both the community-aware and community-unaware persona datasets. While we expected stronger in-group preference when community identity was made explicit, the results show only a minor difference. For example, Community 3 personas chose their own community 41.6\%  of the time when aware, and 42.8\%  when unaware. Similarly, Communities 1 and 2 showed nearly identical in-group selection across both datasets.

We observe that Community 3 is consistently preferred even by out-group personas. This may suggest Community 3 is perceived as most in need, possibly due to its higher SVS of 0.7. These results suggest that community awareness had minimal impact on in-group preference, and that other factors may have influenced decision-making more strongly. Given the high SVS of Community 3, we suspect that personas may have considered broader vulnerability rather than personal affiliation when making prioritization decisions.

%This refers to a persona's preference when its own community is not among the options. We observe that personas from Community 1 highly favor Community 3 (68\%) when selecting outside their own. Community 2 still favors Community 3 (64.1\%),Community 3 is roughly split between Communities 1 and 2 (~51.5\% vs 48.5\%). We see that Community 3 is consistently preferred, even by out-group personas which may suggests that Community 3 may be viewed as most in need, possibly due to their higher SVS (0.7). We conclude that Community 3, despite showing the highest in-group bias, is also the most prioritized by others.

\subsection{Accuracy with Limited Pairwise Comparisons}

We evaluated the possibility of achieving a good global ranking by using the subset of possible pairwise comparisons. We used random sampling to select some samples of the pairwise data, applied chainization and fine-tuning on pseudo pairs to obtain a global rank. Then, we applied Kendall’s $\tau$ correlation between the final inferred ranking and the ranking
obtained from the full dataset.

\begin{figure}[h]
    \centering
    \includegraphics[width=0.5\textwidth]{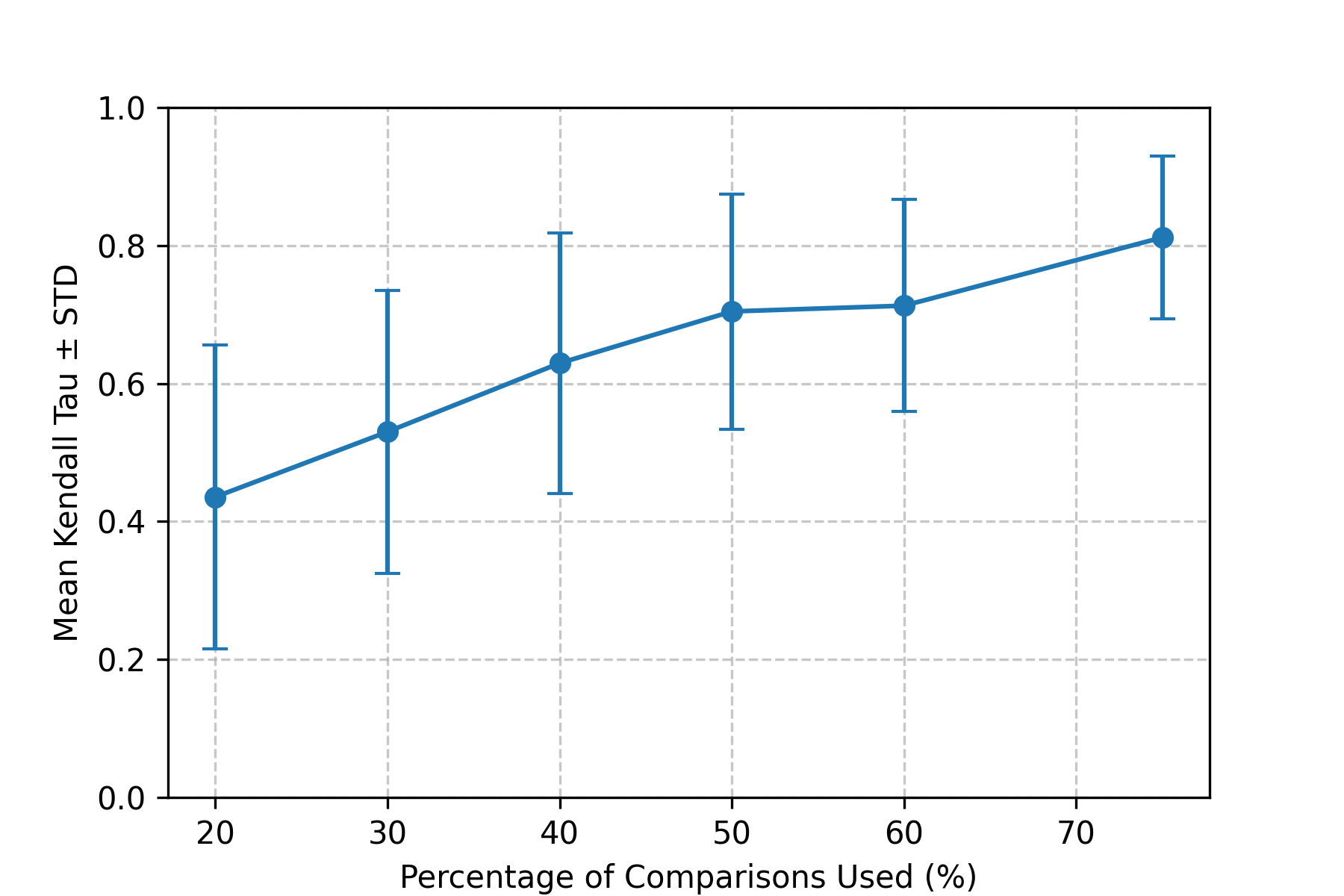}
    \caption{Kendall's $\tau$ correlation versus percentage of comparisons used.}
    \label{fig:kendall_tau_plot}
\end{figure}

Figure~\ref{fig:kendall_tau_plot} shows the trend of mean Kendall's $\tau$ scores across different
percentages of comparisons used. Results are averaged over 100
random trials for each sampling fraction.
We observe that as the percentage of available comparisons increases, the agreement between the predicted and ground-truth rankings improves. We notice that using only 20–30\% of the comparisons achieves a low Kendall's $\tau$ correlation (<0.6), but once the sampled comparisons reach 50\%, the ranking performance improves substantially, reaching a Kendall's $\tau$ above 0.7. These results show that we can recover good global rankings from limited preference data, which offers a practical way to reduce the burden on survey participants while maintaining decision quality.

\begin{figure}[h]
    \centering
    \includegraphics[width=0.5\textwidth]{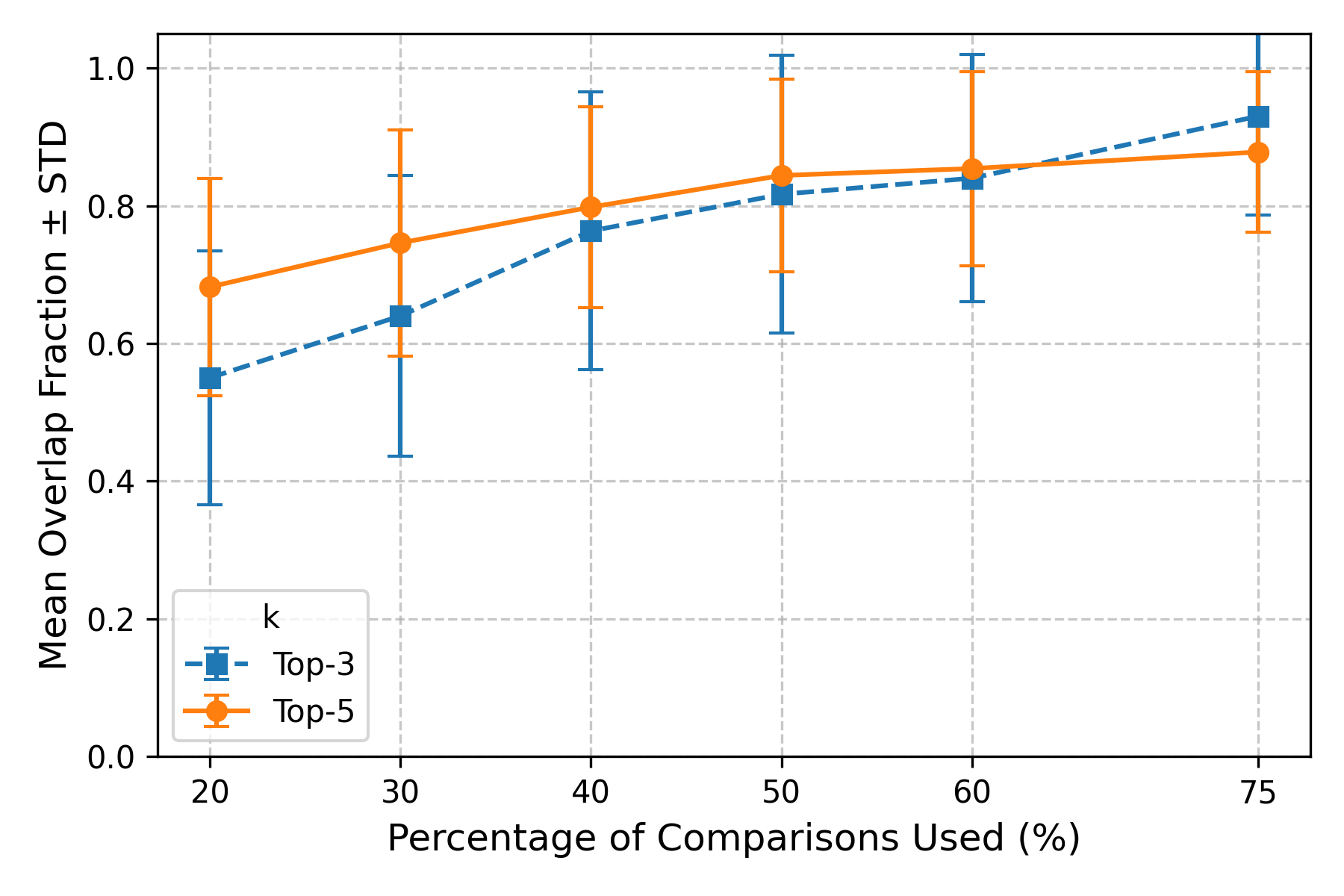}
    \caption{Top-K overlap on Rankings obtained via Random SubSampling and Pseudo-Pairs}
    \label{fig:topk}
\end{figure}

We also present the results of the Top-3 and Top-5 overlaps in Figure~\ref{fig:topk}. The overlap measures how consistently the highest-ranked repair options match the ground truth rankings as we increase the percentage of pairwise comparisons used for training. We observe that the overlap steadily improves as the sample percentage increases. In the range of 40\% to 60\%, the overlaps for both Top-3 and Top-5 remain relatively stable, showing incremental alignment with the ground truth. In fact, the agreement is as high as 80\% . This suggests that a relatively small fraction of the total data is sufficient to capture the most critical repair priorities, which is valuable when collecting survey data is costly or time-constrained.  

\subsection{Sensitivity to Prompt Variations}
\begin{figure}[h]
    \centering
    \includegraphics[width=0.5\textwidth]{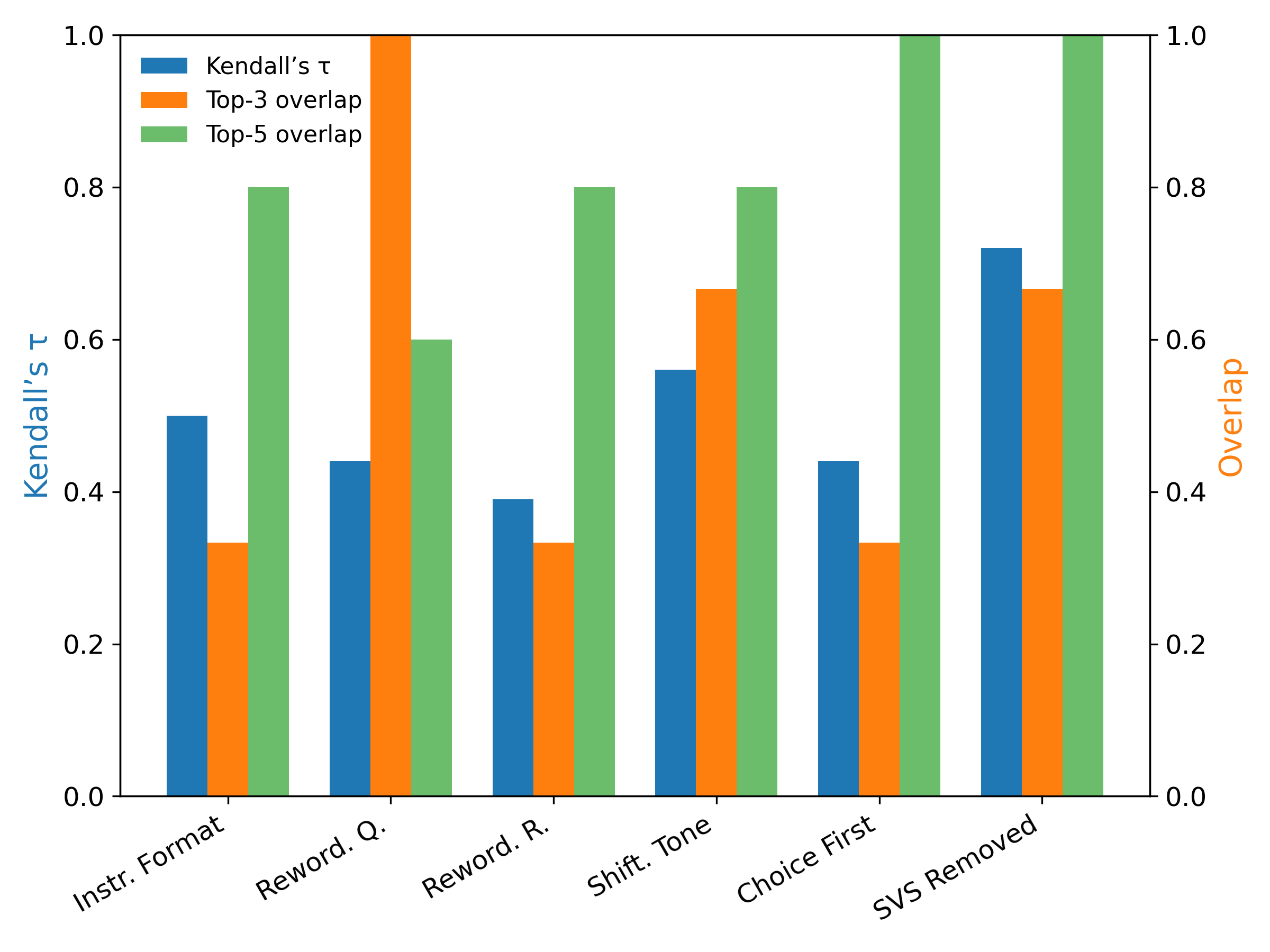}
    \caption{Prompts Variations Overall Agreement with Original using Kendall's $\tau$ and Top-K overlaps.}
    \label{fig:prompt_sensitivity}
\end{figure}

We evaluated six prompt variations against our reference ranking of nine repair options. See Table \ref{tab:prompt_variation_ranking} for each item’s numeric position. Figure~\ref{fig:prompt_sensitivity} summarizes the overall agreement via Kendall’s Tau and Top-$k$ overlaps. We observe that changing instruction format gives $\tau$= 0.50(p=0.075) and rewording the question in the prompts yields $\tau$= 0.44(p=0.12), rewording the reasoning gives $\tau$= 0.39(p=0.18), shifting the tone of prompt yields $\tau$ = 0.56(p=0.045), choice first gives $\tau$ = 0.44(p=0.12) and removing SVS explanation gives $\tau$ = 0.72(p=0.0059). We observe that only the variant where we remove SVS shows high agreement with the original. Using the Top-$K$-Lists overlap, we observe that SVS-removed gives 100\%  Top-5 overlap, rewording the question yields 100\% Top-3 overlap, and other variants show a high level of Top-5 overlap. Our results show that the lower-ranked repairs are more sensitive to prompt changes, and the very top priorities remain stable.

More so, our results indicate that SVS strongly influences persona prioritization and outweighs community affiliation in repair choices. This raises questions about whether the LLM separates vulnerability from affiliation when reasoning, or treats them as interchangeable signals of need. Additionally, our sensitivity analysis shows that the repair rankings remain highly dependent on the prompts provided, with SVS  as a dominant factor.

\begin{table*}[ht]
\centering
\scriptsize
\caption{Global Repair Prioritization Across Prompt Variations}
\label{tab:prompt_variation_ranking}
\begin{tabular}{l|c|c|c|c|c|c}
    \hline
    \makecell{Repair\\Option} & \makecell{Instruction\\Format} & \makecell{Reworded\\Question} & \makecell{Reworded\\Reasoning} & \makecell{Shifting\\Tone} & \makecell{Choice\\First} & \makecell{Without\\SVS} \\
    \hline
    \makecell{Repair Water School\\in Community 3}        & 1 & 1 & 1 & 1 & 3 & 2 \\
    \makecell{Repair Water School\\in Community 1}        & 4 & 2 & 6 & 2 & 5 & 5 \\
    \makecell{Repair Power Residential\\in Community 2}   & 6 & 3 & 5 & 6 & 4 & 1 \\
    \makecell{Repair Power Residential\\in Community 3}   & 3 & 8 & 2 & 5 & 1 & 3 \\
    \makecell{Repair Water Residential\\in Community 3}   & 2 & 7 & 3 & 3 & 2 & 4 \\
    \makecell{Repair Power Residential\\in Community 1}   & 8 & 6 & 9 & 8 & 8 & 6 \\
    \makecell{Repair Water Residential\\in Community 1}   & 7 & 5 & 7 & 7 & 7 & 8 \\
    \makecell{Repair Water Residential\\in Community 2}   & 5 & 4 & 4 & 4 & 6 & 7 \\
    \makecell{Repair Water Commercial\\in Community 2}    & 9 & 9 & 8 & 9 & 9 & 9 \\
    \hline
\end{tabular}
\end{table*}

%We use Kendall’s Tau values to quantify the agreement between each variation and the original ranking. Trespectively.

\section{Discussion}
This study presents an approach to prioritize infrastructure repairs by combining hetero-functional graph modeling with community input generated using large language models (LLMs). Our findings show that combining synthetic persona preferences with repair dependencies can help resolve ambiguity in infrastructure recovery in a scalable and people-centered way.

One of the most important findings in this study is that we can recover accurate global repair rankings using only part of the data. By selecting just 60 percent of the pairwise comparisons and applying chainization with added pseudo-pairs, we achieve high agreement with the full-data rankings. The Kendall Tau correlation goes above 0.7. This is a useful result for disaster planning, where one may lack time or resources to gather full data. Our method reduces the burden on people while still producing reliable prioritization.

We also tested how sensitive our method is to changes in the way we ask questions. When we reworded the prompt, changed its tone, or adjusted the instruction format, the final rankings changed slightly. However, we saw more consistency among higher priority items, while lower ranked repairs were more sensitive to prompt changes. This demonstrates a strong consensus among personas on identifying the most critical tasks.

While our results are promising, there are some limitations. First, the preferences generated by the model may not fully match how real people behave, especially in stressful or emotional situations. Second, the infrastructure model we use is simplified and includes a fixed set of repair tasks. Future research could apply this method to more complex systems with more diverse failure scenarios.

Third, in our approach, we treat all functionalities, regardless of infrastructure type, as part of a unified repair priority list. We acknowledge that this might not necessarily reflect how infrastructure is usually restored in the field. With the exception of certain small communities, water, power, and transportation are usually handled by specialized repair crews and agencies.  Nevertheless, our method can be easily extended to deal with these cases. Moreover, our method can also support broader planning questions, such as where to invest in resilience or how to include community input in upgrade decisions. In such high-level decisions, cross-sector rankings can be helpful, especially when guided by community preferences.

Also, we do not consider social cognitive factors such as measurements of community connectedness and personality traits. Also, we do not compare against classical expert heuristics or existing survey-based methods, or run explicit ablations to isolate the impact of pseudo-pair fine-tuning. Testing on larger, more realistic systems remains unexplored.

Lastly, our experiments with sampling and prompt changes are only a first step. Future work can explore more adaptive strategies that learn which comparisons are most useful as the model is trained.

\section{Conclusion}
Our work presents a new approach to infrastructure repair prioritization by combining hetero-functional graphs, community preferences generated using large language models (LLMs), and a machine learning model that learns from pairwise comparisons. Our method focuses on a common challenge in disaster recovery, which is what to repair first when technical dependencies allow for multiple valid options. By simulating community input through synthetic personas and applying a comparator model with chainization, we are able to produce repair priorities that reflect both system constraints and community preferences.

Our results show that full-data rankings can be approximated using a reduced set of comparisons, lowering the burden of data collection. And we expect that with actual survey data, the selection will likely be more accurate. We also tested how prompt changes affect the outcomes and found that while lower-ranked items shift, top priorities remain stable across variations.

This work opens new directions for integrating synthetic decision-making tools into critical infrastructure resilience frameworks. Based on this, we have identified several future directions for this research.

First, future research should validate these methods with real-world stakeholder input and extend them to more complex systems. We also plan to explore better ways to select informative comparisons and apply the method to a broader range of infrastructure types. Additionally, separating infrastructure types and applying domain-specific repair constraints, incorporating ablation studies, as well as exploring more complex ranking situations, will be important extensions. However, our current unified ranking approach remains valuable for investment planning contexts, where decision-makers must choose across multiple systems.

The next step is to collect survey data and formally test our proof of concept in a real-world setting using actual community data. We also plan to use prior survey data or data from another community and leverage the LLM to generate data for a different community, which may be advantageous given the limited survey response rates in recent years.

An important direction is to address potential biases introduced by the generation of synthetic public viewpoints and responses. We plan to extend this work by validating our approach with real-world survey data that incorporates community demographics, personality characteristics, social cognitive factors, and perception data to enhance the realism and representativeness of the simulated preferences.

We also plan to explore infrastructure consumption nodes as interdependent entities based on resident or household utility demand, allowing for a more integrated modeling of how service disruptions impact communities. This will be a focus in the next phase of our survey-based studies.

In addition, we aim to incorporate community attachment explicitly, as it may significantly influence persona preferences. The community-aware dataset will be expanded and validated using additional metrics of community connectedness in subsequent survey efforts.

Finally, another area for future research is to investigate how repair prioritization rankings might be established pre-disaster, recognizing that real-time or post-disaster decision-making may be challenging. This would allow our framework to support proactive planning scenarios in addition to recovery contexts.

\section*{Acknowledgments}
This research was supported by the U.S. National Science Foundation (NSF) under Grant No. 2148878. We would also like to thank Amin Enderami and Elaina Sutley for their assistance with the toy infrastructure model, and Eugene Vasserman for his valuable feedback on this work.

\appendix\label{appendix}
Here, we provide the original prompts and the different prompt variants. Also, we provide a sample persona and response to the original prompt. Full data, prompt templates, and code are available at \cite{Okeukwu2025_rank}.

\subsection{Prompt Templates}
We show the prompt templates we use to get the repair preferences from LLM-generated personas in our study. The variables in curly brackets(\{\}) are replaced dynamically at runtime with persona attributes and scenario-specific information.

Figure~\ref{fig:orig-prompt} shows the original prompt we use. This prompt format differs from the ``Choice Before Reasoning'' prompt, where the persona makes a choice before reasoning, as seen in this example response~\ref{fig:appendix_response}.

\begin{figure}[H]                     
  \caption{Original Prompt}           
  \label{fig:orig-prompt}
\scriptsize
\begin{framed}
\ttfamily
You are a \{full\_persona\}, and you are aware that a tornado has impacted three communities (1, 2, and 3), causing damage to critical infrastructure. \\

You are familiar with the conditions in all communities, described as follows: \{all\_community\_descriptions\}. Social Vulnerability Scores (SVS) range from 0 (least vulnerable) to 1 (most vulnerable), with higher values indicating greater vulnerability.
 \\

Consider the situation, taking into account any details you find relevant. \\

Question: Which should be repaired first:`\{option1\}' or `\{option2\}'? \\

Before choosing between `\{option1\}' and `\{option2\}', first explain your reasoning. \\

Provide your answer in the following format: \\
Reasoning: [Your one-sentence explanation] \\
Priority Choice: [Your chosen option]
\end{framed}
\normalsize
\end{figure}

Figure~\ref{fig:instruction-prompt} shows the prompt for changing the instruction format.

\begin{figure}[H]                     
  \caption{Changing Instruction Format}           
  \label{fig:instruction-prompt}
\scriptsize
\begin{framed}
\ttfamily
Step 1: You are a \{full\_persona\}, and you are aware that a tornado has impacted three communities (1, 2, and 3), causing damage to critical infrastructure. \\

Step 2: You are familiar with the conditions in all communities, described as follows: \{all\_community\_descriptions\}. Social Vulnerability Scores (SVS) range from 0 (least vulnerable) to 1 (most vulnerable), with higher values indicating greater vulnerability.
 \\

Step 3: Consider the situation, taking into account any details you find relevant. \\

Step 4: Question: Which should be repaired first:`\{option1\}' or `\{option2\}'? \\

Step 5: Before choosing between `\{option1\}' and `\{option2\}', first explain your reasoning. \\

Step 6: Provide your answer in the following format: \\
Reasoning: [Your one-sentence explanation] \\
Priority Choice: [Your chosen option]
\end{framed}
\normalsize
\end{figure}

Figure~\ref{fig:shifting-prompt} shows the prompt for shifting tone.

\begin{figure}[H]                     
  \caption{Shifting Tone Prompt}           
  \label{fig:shifting-prompt}
\scriptsize
\begin{framed}
\ttfamily
You are a \{full\_persona\}, and you are aware that a tornado has impacted three communities (1, 2, and 3), causing damage to critical infrastructure. \\

You are familiar with the conditions in all communities, described as follows: \{all\_community\_descriptions\}. Social Vulnerability Scores (SVS) range from 0 (least vulnerable) to 1 (most vulnerable), with higher values indicating greater vulnerability.
 \\
 
Keep in mind the emotional and practical consequences of the outage for residents living in these communities as you evaluate what should be repaired first.\\

Consider the situation, taking into account any details you find relevant. \\

Question: Which should be repaired first:`\{option1\}' or `\{option2\}'? \\

Before choosing between `\{option1\}' and `\{option2\}', first explain your reasoning. \\

Provide your answer in the following format: \\
Reasoning: [Your one-sentence explanation] \\
Priority Choice: [Your chosen option]
\end{framed}
\normalsize
\end{figure}

Figure~\ref{fig:question-prompt} shows the prompt for rewording the question.

\begin{figure}[H]                     
  \caption{Rewording Question}           
  \label{fig:question-prompt}
\scriptsize
\begin{framed}
\ttfamily
You are a \{full\_persona\}, and you are aware that a tornado has impacted three communities (1, 2, and 3), causing damage to critical infrastructure. \\

You are familiar with the conditions in all communities, described as follows: \{all\_community\_descriptions\}. Social Vulnerability Scores (SVS) range from 0 (least vulnerable) to 1 (most vulnerable), with higher values indicating greater vulnerability.\\

Consider the situation, taking into account any details you find relevant. \\

Question: Between the two options, `\{option1\}' and `\{option2\}', which do you deem most critical for immediate repair? \\
                   
Before choosing between `\{option1\}' and `\{option2\}', first explain your reasoning. \\

Provide your answer in the following format: \\
Reasoning: [Your one-sentence explanation] \\
Priority Choice: [Your chosen option]
\end{framed}
\normalsize
\end{figure}

Figure~\ref{fig:reason-prompt} shows the prompt for rewording the reasoning.
\begin{figure}[H]                     
  \caption{Rewording Reasoning}           
  \label{fig:reason-prompt}
\scriptsize
\begin{framed}
\ttfamily
You are a \{full\_persona\}, and you are aware that a tornado has impacted three communities (1, 2, and 3), causing damage to critical infrastructure. \\

You are familiar with the conditions in all communities, described as follows: \{all\_community\_descriptions\}. Social Vulnerability Scores (SVS) range from 0 (least vulnerable) to 1 (most vulnerable), with higher values indicating greater vulnerability.
 \\

Consider the situation, taking into account any details you find relevant. \\

Question: Which should be repaired first:`\{option1\}' or `\{option2\}'? \\

Before choosing between `\{option1\}' and `\{option2\}', first explain your reasoning. \\

Provide your answer in the following format: \\
Reasoning: [Explain your selection with a concise rationale, highlighting relevant factors influencing your decision.] \\
Priority Choice: [Your chosen option]
\end{framed}
\normalsize
\end{figure}

\subsection{Sample Persona and Response}
Figure~\ref{fig:persona_example} is an example persona created by the LLM and used in the simulation. Figure~\ref{fig:appendix_response} is an example response to the pairwise prioritization question between choosing Power in Residential in Community 1 and Power in Residential in Community 3.

\captionsetup{type=figure}

\begin{lstlisting}
  "persona_id": "P002",
  "age": 30,
  "gender": "Male",
  "race_ethnicity": "Hispanic",
  "marital_status": "Married",
  "citizenship_status": "Naturalized Immigrant",
  "birth_place": "Mexico",
  "year_of_immigration": null,
  "primary_language": "Spanish",
  "english_proficiency": "Fluent",
  "education_level": "Master's Degree",
  "school_enrolment": "No",
  "employment_status": "Employed",
  "occupation": "Software Engineer",
  "industry_of_employment": "Technology",
  "class_of_worker": "Private Sector",
  "work_hours_per_week": 40,
  "income_bracket": "$50K-$75K",
  "poverty_status": "Above Poverty Line",
  "government_assistance": "None",
  "household_relationship": "Married",
  "household_size": 2,
  "number_of_children": 2,
  "responsible_for_grandchildren": false,
  "home_ownership": "Own",
  "housing_type": "Single-Family Home",
  "utilities_access": "Full access to utilities",
  "utilities_housing_costs": "$1500\_per\_month",
  "transportation_mode": "Drives",
  "commute_duration": 30,
  "work_from_home": false,
  "health_insurance": "Employer-provided",
  "disability_status": "No",
  "disability_type": null,
  "military_service": "None",
  "military_service_period": null,
  "disability_rating": null,
  "internet_access": "Broadband",
  "device_access": "Laptop",
  "recent_birth": true,
  "moved_in_last_year": false,
  "previous_residence": "Same city",
  "disaster_experience": "Experienced an earthquake in 2018",
  "community_description": "In Community 1: ... (text truncated for brevity)",
  "geographical_class": "Suburban"
\end{lstlisting}

\captionof{figure}{Full specification of one synthetic persona used in the simulation}
\label{fig:persona_example}

\begin{figure}[H]                     
  \caption{Sample Response}           
  \label{fig:appendix_response}
\scriptsize
\begin{framed}
\ttfamily
\texttt{Question:} \\
\texttt{Which should be repaired first: Power Residential in Community 1} \\
\texttt{or Power Residential in Community 3?} \\

\vspace{2mm}
\texttt{Response:} \\
\texttt{Priority Choice: Repair Power Residential in Community 3} \\
\texttt{Reasoning: Community 3 has a higher social vulnerability score} \\
\texttt{of 7.0/10 compared to Community 1's 2.0/10, indicating a greater} \\
\texttt{need for immediate assistance to ensure safety in the more} \\
\texttt{vulnerable urban area.}
\end{framed}
\normalsize
\end{figure}

%\begin{thebibliography}{10}

%\bibitem{c1}

%\end{thebibliography}

%\bibliographystyle{IEEEtran}
%\bibliography{ref}
\input{ref.bbl}

\end{document}